\documentclass[aps,prl,twocolumn,showpacs]{revtex4}

\usepackage{graphicx}
\usepackage{dcolumn}
\usepackage{amsmath}
\usepackage{amssymb}
\usepackage{wasysym}
\usepackage{color}

\usepackage{mathtools}

\DeclarePairedDelimiter\floor{\lfloor}{\rfloor}

\begin{document}

\title{Change in sign of the Hall coefficient from Fermi surface curvature in underdoped high $T_{\rm c}$  copper oxide superconductors}

\author{N.~Harrison$^1$, S.~E.~Sebastian$^2$
}

\affiliation{$^1$Mail~Stop~E536,~Los~Alamos~National Labs.,Los~Alamos,~NM~ 87545\\
$^2$Cavendish Laboratory, Cambridge University, JJ Thomson Avenue, Cambridge CB3~OHE, U.K
}
\date{\today}

\begin{abstract}
It has recently been proposed that the Fermi surface of underdoped high $T_{\rm c}$ copper oxide materials within the charge-ordered regime consists of a diamond-shaped electron pocket constructed from arcs connected at vertices. We show here that on modeling the in-plane magnetotransport of such a Fermi surface using the Shockley-Chambers tube integral approach and a uniform scattering time, several key features of the normal state in-plane transport of the underdoped copper oxide systems can be understood. These include the sign reversal in the Hall coefficient, the positive magnetoresistance and magnetic quantum oscillations in the Hall coefficient. 
\end{abstract}
\maketitle

\begin{figure}
\centering 
\includegraphics*[width=.45\textwidth]{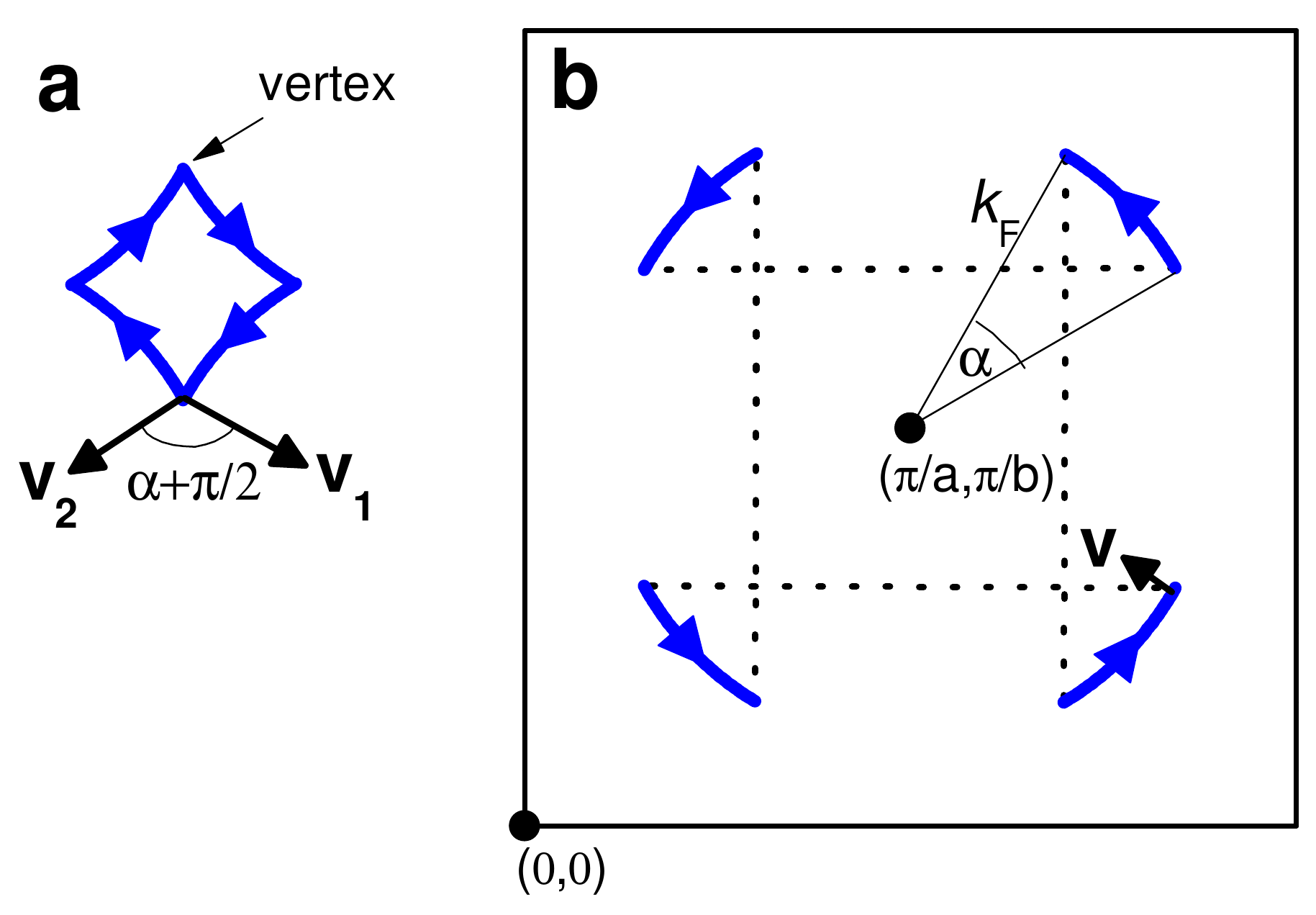}
\caption{({\bf a}), Schematic diamond-shaped electron pocket~\cite{harrison1,sebastian1,sebastian9,allais1,zhang1,tabis1,doiron2,senthil1}, with blue arrows indicating the direction of cyclotron motion and ${\bf v}_1$ and ${\bf v}_2$ indicating the Fermi velocity direction. ({\bf b}), Schematic showing how the electron pocket is produced by connecting `arcs' of a larger hole Fermi surface, with $\alpha$ being the angle subtended by the arc and the dotted lines indicating how they are connected.}
\label{construction}
\end{figure}

Aside from their extraordinarily high superconducting transition temperatures, the layered copper oxide materials are known for their unusual normal state transport properties, which include sign reversals in the Hall coefficient at low temperatures~\cite{harris1,adachi,lin1,leboeuf1,doiron3}. Recent quantum oscillation experiments that reveal small Fermi pockets of area $\approx$~2~\% of the Brillouin zone in the underdoped hole-doped cuprate superconductors~\cite{doiron1,barisic1,yelland1,bangura1} rather than a large Fermi surface occupying $\approx$~40~\% of the Brillouin zone predicted by band structure~\cite{andersen1}, have led to proposals associating the sign reversal in Hall coefficient with Fermi surface reconstruction~\cite{millis1,chakravarty1,yao1,harrison1}. Complementary measurements, which include the Hall effect~\cite{leboeuf1,doiron3}, nuclear magnetic resonance~\cite{wu1}, x-ray scattering~\cite{ghiringhelli1,chang1}, magnetic quantum oscillations~\cite{doiron1,barisic1,yelland1,bangura1} and others, point to a form of charge ordering that breaks the large hole-like Fermi surface predicted by band structure in underdoped YBa$_2$Cu$_3$O$_{6+x}$, YBa$_2$Cu$_4$O$_8$ and HgBa$_2$CuO$_{4+\delta}$ into small electron pockets. Biaxial charge ordering, in particular, has been proposed to yield a reconstructed Fermi surface consisting of a diamond-shaped electron pocket located in the nodal region of the original Brillouin zone~\cite{harrison1,sebastian1,sebastian9,allais1,zhang1,tabis1,doiron2,senthil1}, resembling that in Fig.~\ref{construction}.

While the observed sign reversals in the Hall coefficient $R_{\rm H}$ from positive to negative at low temperatures occur at tens of kelvin, the transitions into the low temperature short-range charge ordered phases are reported to begin at much higher temperatures~\cite{leboeuf2,ghiringhelli1,chang1}. The overall form of the crossover from positive to negative is therefore suggested to depend on the specific Fermi surface topology. Such a crossover can be modeled by circular pockets of electrons and holes with very different mobilities~\cite{chakravarty1,rourke1,doiron2}. Alternatively, the sign reversal could originate from a form of $R_{\rm H}$ that is driven by the curvature of the Fermi surface~\cite{banik1}. It is intriguing to consider the latter possibility given the similarity of the proposed diamond-shaped electron pocket~\cite{harrison1,sebastian1,sebastian9,allais1,zhang1,tabis1,doiron2,senthil1} to the Fermi surface cross-sections originally considered by Banik and Overhauser~\cite{banik1}. 

\begin{figure}
\centering 
\includegraphics*[width=.45\textwidth]{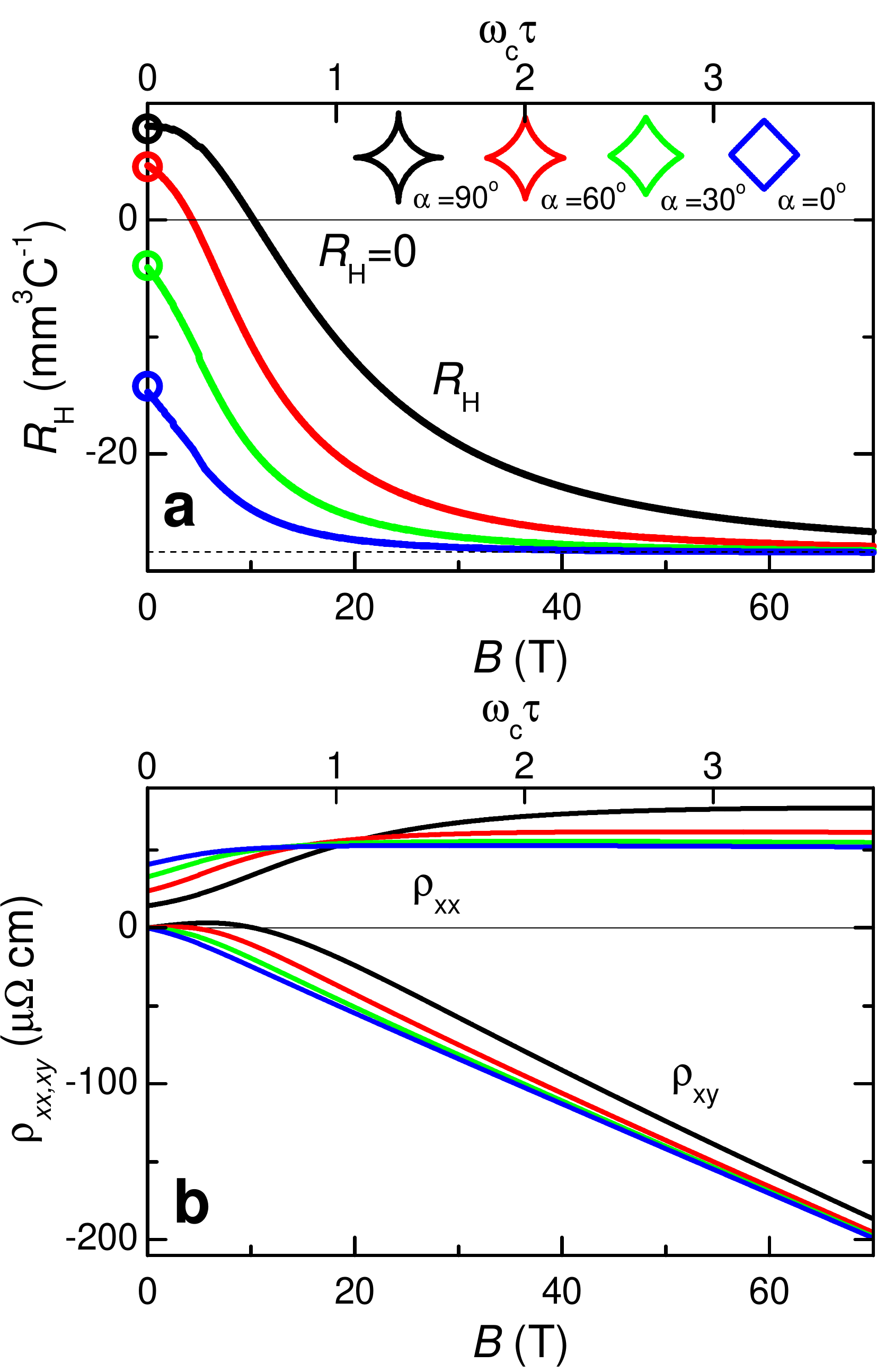}
\caption{({\bf a}), $R_{\rm H}$ in the weak magnetic field limit (colored circles) according to Equation~(\ref{diamondhall}) and at arbitrary magnetic fields (colored lines) according to Equations~(\ref{stronghall}) and (\ref{tubeintegral}). The different colours indicate different values of $\alpha$, the corresponding pocket shapes for which are shown in the top-right-hand corner. The horizontal solid line indicates $R_{\rm H}=0$ while the horizontal dotted line indicates the high magnetic field limit of $R_{\rm H}$. ({\bf b}), $\rho_{xy}=\sigma_{xy}/(\sigma_{xy}^2+\sigma_{xx}^2)$ and $\rho_{xx}=\sigma_{xx}/(\sigma_{xy}^2+\sigma_{xx}^2)$ according to Equation~(\ref{tubeintegral}) calculated for the same values of $\alpha$. For two copper oxide planes per unit cell, these values must be halved.
}
\label{hall}
\end{figure}

The effect of the Fermi surface curvature on the sign of $R_{\rm H}$ depends on the variation of the scattering time $\tau$ and Fermi velocity ${\bf v}$ around the orbit. In this paper, we show that by using the Shockley-Chambers tube integral approach~\cite{shockley1,chambers1} for the case of a single uniform scattering time, the magnetic field and temperature-dependence of $R_{\rm H}$ yielded by a nodal diamond-shaped pocket changes in sign, and exhibits a form closely resembling that measured in YBa$_2$Cu$_3$O$_{6+x}$, YBa$_2$Cu$_4$O$_8$ and HgBa$_2$CuO$_{4+\delta}$~\cite{leboeuf1,doiron3} (see Fig.~\ref{hall}). We show these results to span both the strong magnetic field limit and the weak magnetic field Jones-Zener limit~\cite{jones1}. We further show that a diamond-shaped pocket can explain the presence of in-plane magnetoresistance and the appearance of large amplitude quantum oscillations in the Hall coefficient~\cite{doiron1}, without the need to invoke multiple Fermi surface pockets. 

For a determination of the Hall coefficient 
\begin{equation}\label{stronghall}
R_{\rm H}=\frac{1}{B}~\frac{\sigma_{xy}}{\sigma_{xy}^2+\sigma_{xx}^2},
\end{equation}
in a magnetic field of arbitrary strength, we calculate the diagonal ($\sigma_{xx}$) and off-diagonal ($\sigma_{xy}$) conductivity components using
\begin{equation}\label{tubeintegral}
\sigma_{x\beta}=-\frac{e^3B}{2\pi^2\hbar^2}\int_0^T\Big(\int_0^\infty v_x(t)e^{-\frac{t^\prime}{\tau(t)}}v_\beta(t+t^\prime){\rm d}t^\prime\Big){\rm d}t,
\end{equation}
which is the Shockley-Chambers tube integral~\cite{chambers1,shockley1} expressed as a function of time $t$, with the subscript $\beta$ referring to $x$ or $y$. We neglect the dependence of the velocity ${\bf v}$ on energy, but consider the variation of its components around the diamond-shaped electron pocket orbit in time. These are given by 
$v_x=v_{\rm F}\cos\big(\big((\omega t\mod{\frac{\pi}{2}})-\frac{\pi}{4}\big)\frac{2\alpha}{\pi}+\frac{\pi}{4}-\floor[\big]{\frac{2\omega t}{\pi}}\pi\big)$ and $v_y=v_{\rm F}\sin\big(\big((\omega t\mod{\frac{\pi}{2}})-\frac{\pi}{4}\big)\frac{2\alpha}{\pi}+\frac{\pi}{4}-\floor[\big]{\frac{2\omega t}{\pi}}\pi\big)$, where $\omega=eB/m$ is the cyclotron frequency of the diamond-shaped orbit. Figure~\ref{hall} shows $R_{\rm H}$ calculated for four different values of the angle $\alpha$ subtended by the concave sides of the pocket (shown schematically in Fig.~\ref{construction}b). For $\alpha=$~90$^\circ$, our results reproduce the findings of Banik and Overhauser~\cite{banik1}. 

As expected, $R_{\rm H}$ saturates at $-\frac{1}{en}$ (dotted line in Fig.~\ref{hall}b) in the strong magnetic field limit, where $n=\frac{A_k}{2\pi^2c}$ is the carrier density and $c\approx$~11.68~\AA~ is the interlayer lattice constant. We use Onsager's relation $F=\frac{\hbar}{2\pi e}A_k$~\cite{shoenberg1}
to obtain the Fermi surface cross-sectional area $A_k$ in momentum-space for a pocket of frequency $F\approx$~530~T~\cite{doiron1}.

To understand $R_{\rm H}$ in the weak magnetic field limit, we turn to the Jones-Zener solution to the Boltzmann transport equation~\cite{jones1}. Ong~\cite{ong1} has shown that the Hall coefficient
\begin{equation}\label{ong}
R_{\rm H}=(2\pi)^2cA_l/e\pi(\bar{|{\bf l}|}S)^2
\end{equation}
can be re-expressed in terms of the Stokes area $A_l=(\frac{\bf B}{B})\cdot\int\frac{1}{2}{\rm d}{\bf l}\times{\bf l}$ swept out by the `scattering path length' vector ${\bf l}=\tau{\bf v}$ on moving around the Fermi surface perimeter.
Here, $\bar{|{\bf l}|}$ is the orbital average of the scattering path length, while $B$ is the magnitude of the magnetic field ${\bf B}$. We define the perimeter of the diamond-shaped Fermi surface $S=4\alpha k_{\rm F}$ in terms of the effective radius $k_{\rm F}$ of the large hole-like circular Fermi surface from which the arcs originate in Fig.~\ref{construction}b.
The angular variation of ${\bf v}$ at the Fermi surface vertices requires special care in the Jones-Zener method~\cite{ong1}, unlike in the Shockley-Chambers tube integral treatment~\cite{shockley1,chambers1}, where contributions from the vertices are implicitly included.

Vertices occur when nearly free electrons (or holes in the case of the cuprates) are perturbed by a weak periodic potential. As a consequence, Fermi surface sections with velocities in different directions intersect owing to a translation of the Fermi surface by a new lattice vector ${\bf K}$. The origin of such Fermi surface translation arises from a charge ordering potential in the underdoped cuprates~\cite{ghiringhelli1,chang1}, while in elemental systems like Al, it originates from the crystalline lattice potential~\cite{banik1}.

\begin{figure}
\centering 
\includegraphics*[width=.45\textwidth]{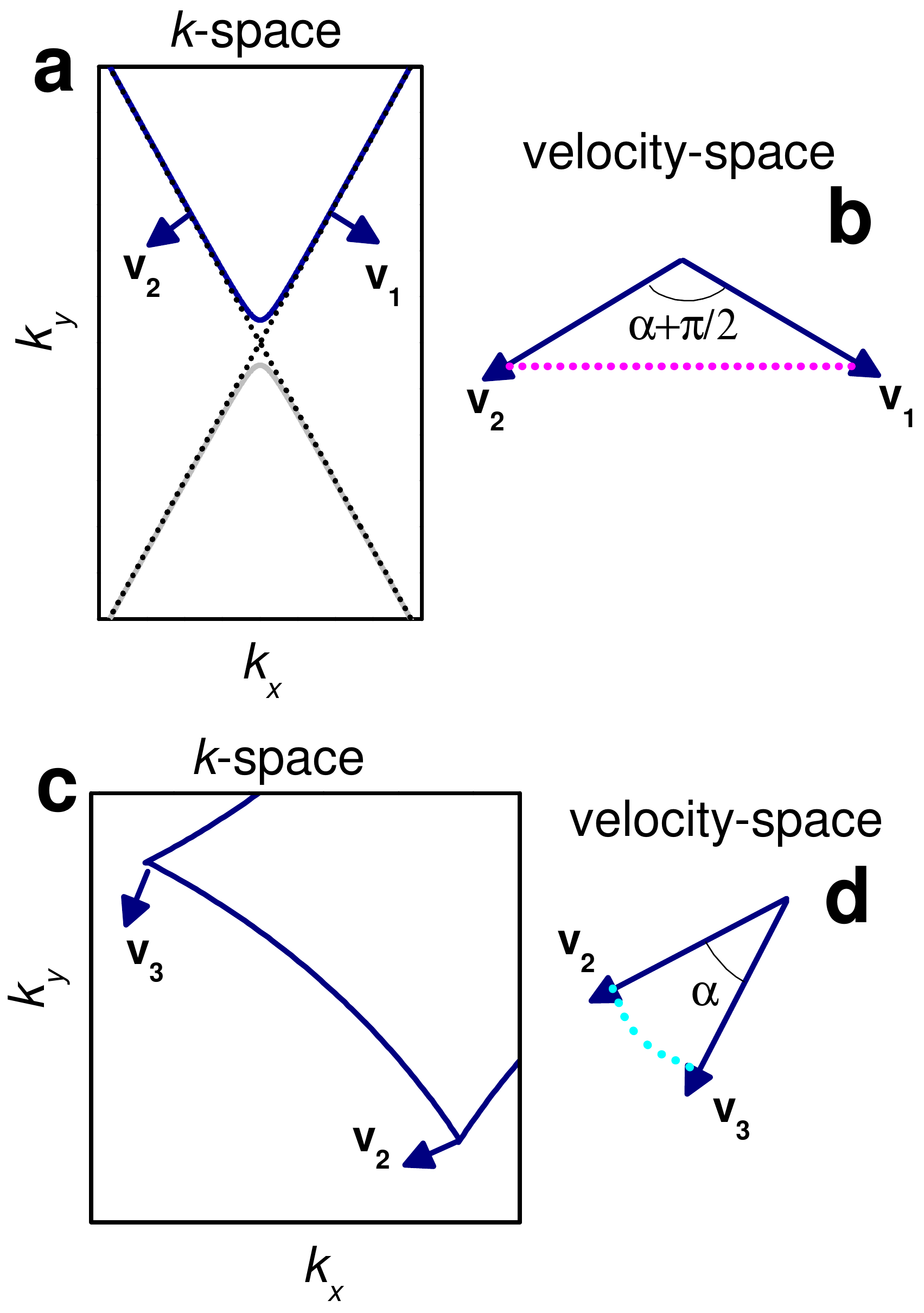}
\caption{{\bf a}, Solid lines showing the reconstructed Fermi surface in the vicinity of a vertex according to Equation (\ref{gapped}). Dotted lines indicate the Fermi surface in the absence of hybridization. ${\bf v}_1$ and ${\bf v}_2$ are velocities before and after a quasiparticle traverses the vertex. ({\bf b}) Schematic of the triangular area swept out by the velocity vector at a vertex, bounded by the arrows and the magenta dotted line. ({\bf c}), Schematic showing the change in velocity along the concave arc-like portion of a Fermi surface orbit. ({\bf d}), Schematic of the area (in this case a sector of a circle) swept out by the velocity vector along the Fermi surface arc, bounded by the arrows and the cyan dotted line.}
\label{diagram}
\end{figure}

We model such a vertex of the diamond-shaped pocket in Fig.~\ref{construction}a by making linear approximations to the electronic dispersion in the vicinity of a point of intersection between two arcs. Upon choosing the point of intersection to be located at the origin in our $k$-space coordinates, the untranslated and translated dispersions acquire the simple forms $\varepsilon_{\bf k}=\hbar v_{\rm F}\big(-k_x\sin(\frac{\alpha}{2}+\frac{\pi}{4})-k_y\cos(\frac{\alpha}{2}+\frac{\pi}{4})\big)$ and $\varepsilon_{{\bf k}+{\bf K}}=\hbar v_{\rm F}\big(k_x\sin(\frac{\alpha}{2}+\frac{\pi}{4})-k_y\cos(\frac{\alpha}{2}+\frac{\pi}{4})\big)$, respectively, where $v_{\rm F}$ is the constant magnitude of the Fermi velocity on the arcs. The Fermi surfaces at $\varepsilon_{\bf k}=0$ and $\varepsilon_{{\bf k}+{\bf K}}=0$ are represented by dotted lines in Fig.~\ref{diagram}a. A finite amplitude periodic potential opens up a small hybridization gap of magnitude $2\Delta$ between these dispersions, leading to reconstructed dispersions of the form
\begin{eqnarray}\label{gapped}
\varepsilon_{\bf k}^\pm=\hbar v_{\rm F}\bigg(-k_y\cos\big(\frac{\alpha}{2}+\frac{\pi}{4}\big)\hspace{3cm}\nonumber\\
\pm\sqrt{k_x^2\sin^2\big(\frac{\alpha}{2}+\frac{\pi}{4}\big)+\frac{\Delta^2}{\hbar^2v^2_{\rm F}}}~~\bigg)
\end{eqnarray}
whose modified Fermi surfaces at $\varepsilon^\pm_{\bf k}=0$ are indicated by solid lines in Fig.~\ref{diagram}a. Using ${\bf v}=\frac{1}{\hbar}\nabla{\varepsilon}_{\bf k}^+$, we obtain \begin{equation}
{\bf v}=v_{\rm F}\Bigg[\frac{k_x\sin^2\big(\frac{\alpha}{2}+\frac{\pi}{4}\big)}{\sqrt{k_x^2\sin^2\big(\frac{\alpha}{2}+\frac{\pi}{4}\big)+\frac{\Delta^2}{\hbar^2v^2_{\rm F}}}},-\cos\big(\frac{\alpha}{2}+\frac{\pi}{4}\big)\Bigg]
\end{equation}
for the velocity. The $x$ component of the velocity changes from $-v_{\rm F}\sin\big(\frac{\alpha}{2}+\frac{\pi}{4}\big)$ to $+v_{\rm F}\sin\big(\frac{\alpha}{2}+\frac{\pi}{4}\big)$ on varying $k_x$ from $-\infty$ to $+\infty$, while the $y$ component remains at a constant value of $-v_{\rm F}\cos\big(\frac{\alpha}{2}+\frac{\pi}{4}\big)$. 

To compute the total scattering path area ($A_l$) swept out, contributions are added from both the vertex regions, and the arc regions.  The vertex contribution to the scattering path area is identified as the triangular area swept out between ${\bf v}_1$ and ${\bf v}_2$ in Fig.~\ref{diagram}b multiplied by $\tau^2$, yielding $A_{\rm v}=-\frac{1}{2}(\tau v_{\rm F})^2\cos\alpha$ in the limit of small $\Delta$. The arc contribution to the scattering path area (e.g. Fig.~\ref{diagram}c) is identified as the sector of a circle swept out between ${\bf v}_2$ and ${\bf v}_3$ in Fig.~\ref{diagram}d multiplied by $\tau^2$, yielding  $A_{\rm a}=\frac{1}{2}(\tau v_{\rm F})^2\alpha$. On summing the contributions to $A_l$ from 4 vertices and 4 arcs, we arrive at 
\begin{equation}\label{diamondhall}
R_{\rm H}=\frac{1}{en}\frac{(\alpha-\cos\alpha)(\sin\alpha+1-\cos\alpha-\alpha)}{2\alpha^2},
\end{equation}
where $n=2Nk_{\rm F}^2(\sin\alpha+1-\cos\alpha-\alpha)/\pi^2c$ is the carrier density of the diamond-shaped pocket. In Fig.~\ref{hall}, the weak magnetic field limit Hall coefficient given by Equation (\ref{diamondhall}) is seen to yield identical results to those obtained using the Shockley-Chambers tube integral when $\tau$ is taken to be uniform. 

The sign of the Hall coefficient is determined by the net of the opposing vertex and arc contributions, which is reflected in the first term within parenthesis in the numerator of Equation~(\ref{diamondhall}). For uniform $\tau$, the condition for obtaining a net positive Hall coefficient in this weak magnetic field limit is $|A_{\rm a}|>|A_{\rm v}|$, corresponding to $\alpha>\cos\alpha$ (i.e. $\alpha>$~42.3$^\circ$) in Fig.~\ref{construction}. This therefore sets the condition for a sign reversal to occur on increasing $\omega\tau$ in Fig.~\ref{hall}.

\begin{figure}
\centering 
\includegraphics*[width=.45\textwidth]{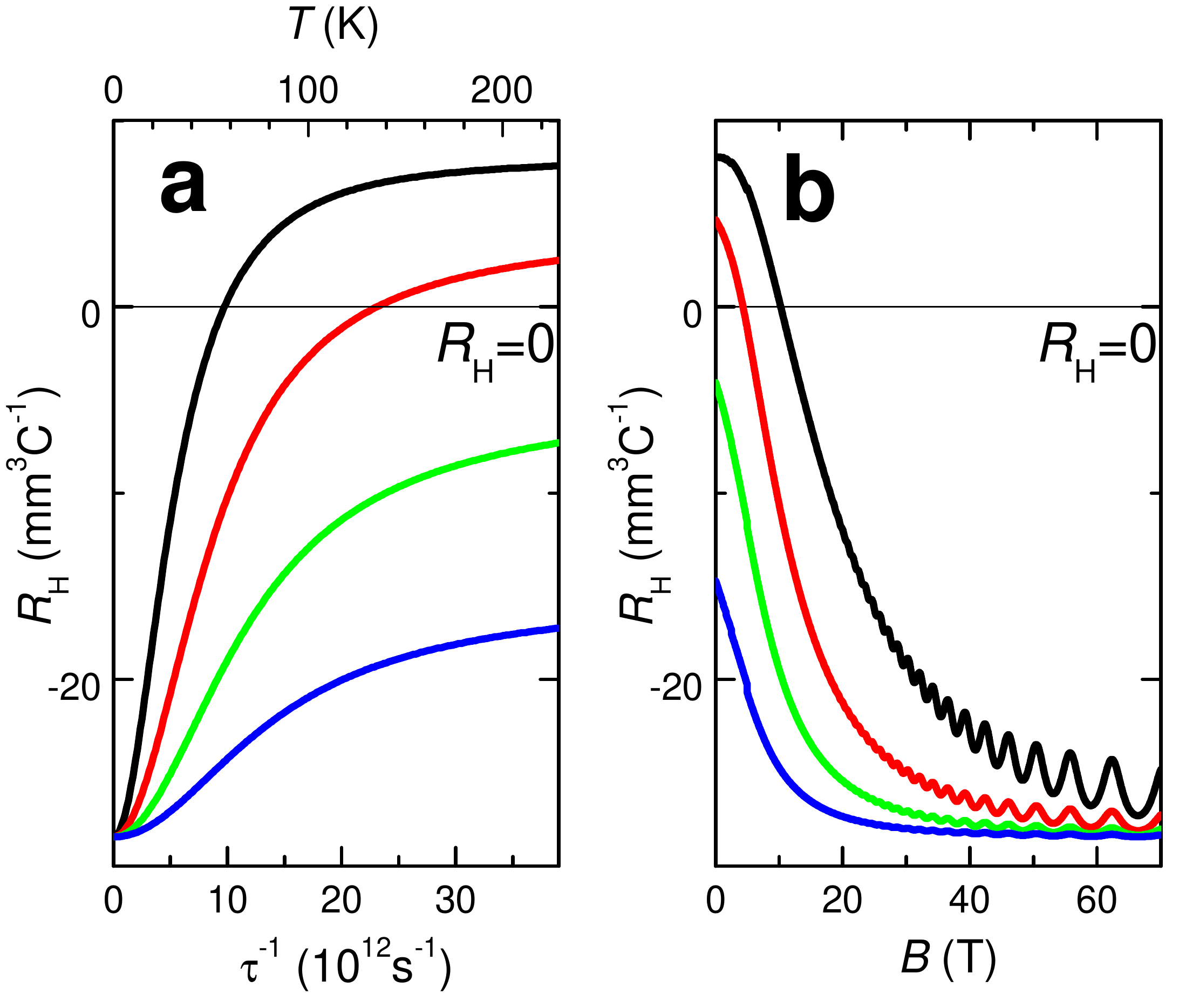}
\caption{({\bf a}), $R_{\rm H}$ versus $\tau^{-1}$ (lower axis) at a constant magnetic field of 50~T. The upper axis shows the corresponding `Dingle temperature' $T=1/2\pi\tau k_{\rm B}$~\cite{shoenberg1}. ({\bf b}), $R_{\rm H}$ versus $B$ calculated using a quantum oscillatory $\tilde{\tau}^{-1}\approx(1+2\cos(\frac{2\pi F}{B}-\pi)e^{-\frac{\pi}{\omega\tau}})\tau^{-1}$, using $\tau=$~0.5~$\times$~10$^{-12}$~\cite{sebastian1} (estimated from global fits to experimental data).}
\label{oscillations}
\end{figure}

On considering only the simplest case of a uniform $\tau$, key features of the Hall coefficient and magnetoresistance data in the underdoped systems YBa$_2$Cu$_3$O$_{6+x}$, YBa$_2$Cu$_4$O$_8$ and HgBa$_2$CuO$_{4+\delta}$~\cite{leboeuf1,doiron3} naturally arise from a diamond-shaped pocket with $\alpha>$~42.3$^\circ$ in Fig.~\ref{construction}. {\it First}, the calculated $R_{\rm H}$ in Fig.~\ref{hall}a changes sign from positive to negative on increasing $\omega\tau$ from 0 to $\approx$~1 (upper axis) in a manner that is qualitatively similar to that seen experimentally in the non-superconducting regime of magnetic field and temperature. The lower axis in Fig.~\ref{hall}a represents the corresponding magnetic field for $\tau=$~0.5~$\times$~10$^{-12}$~s~\cite{sebastian1} and an effective mass of $m=$~1.6~$m_{\rm e}$ (where $m_{\rm e}$ is the free electron mass), which are the parameters obtained from YBa$_2$Cu$_3$O$_{6+x}$ quantum oscillation measurements~\cite{sebastian1}. {\it Second}, the Hall sign reversal is accompanied by magnetoresistance in $\rho_{xx}$ in Fig.~\ref{hall}b, which resembles the observed magnetoresistance in transport experiments~\cite{leboeuf1,doiron3}. In the calculations (see Fig.~\ref{hall}b), the magnetoresistance is found to increase with $\alpha$. {\it Third}, The calculated form of $R_{\rm H}$ as a function of $\tau^{-1}$ in constant magnetic field in Fig.~\ref{oscillations}a, is found to be qualitatively consistent with temperature-dependent measurements of $R_{\rm H}$~\cite{leboeuf1,doiron3}. The upper axis shows the corresponding `Dingle temperature~\cite{shoenberg1},' which approximates the true temperature in the case where $\tau^{-1}$ linearly increases with temperature.
{\it Finally}, the diamond-shaped Fermi surface pocket provides a natural explanation for the observation of quantum oscillations in the Hall coefficient of YBa$_2$Cu$_3$O$_{6+x}$~\cite{doiron1,leboeuf1}, without the need to invoke multiple Fermi surface pockets~\cite{doiron2,sebastian9,chakravarty1}. Quantum oscillations in the electronic density-of-states lead directly via Fermi's golden rule~\cite{pippard1,kikugawa1} to an oscillatory scattering rate $\tilde{\tau}^{-1}$. Whereas $\tilde{\tau}$ factorizes out of the semiclassical expression for $R_{\rm H}$ in the case of a circular Fermi surface pocket~\cite{kikugawa1}, this is no longer the case for a Fermi surface consisting of concave sections~\cite{banik1}. On substituting $\tilde{\tau}^{-1}$ in place of $\tau^{-1}$ in the calculations, we find quantum oscillations in $R_{\rm H}$ that are a significant fraction of $R_{\rm H}$ in Fig.~\ref{oscillations}b when $\alpha$ is large.

Our calculations show the observed sign reversal in the Hall effect, quantum oscillations in the Hall coefficient, and magnetoresistance in transport occurs for a single diamond-shaped pocket of sufficiently concave curvature, properties of which are also in better agreement with complementary measurements such as heat capacity~\cite{riggs1} than various multiple Fermi surface pocket scenarios~\cite{doiron2,chakravarty1,rourke1}.

Our conclusions remain intact if $\tau^{-1}$~\cite{narduzzo1,analytis1} (or alternatively $m$~\cite{senthil1}) increases gradually on moving from the middle of the arcs (i.e. the nodal direction) towards the direction of the antinodes, as reported for the large unreconstructed Fermi surfaces of La$_{2-x}$Sr$_x$CuO$_4$ and Tl$_2$Ba$_2$CuO$_{6+\delta}$~\cite{narduzzo1,analytis1}. The resulting reduction in the negative contribution to $R_{\rm H}$ from the vertices relative to the positive one from the arc regions, yields a sign reversal for smaller values of $\alpha$ than in the uniform $\tau^{-1}$ scenario. A different scenario by which the scattering length retains a constant magnitude while rounding the vertices has been suggested in ref.~\cite{ong1}, yielding a magnetic field-independent sign of $R_{\rm H}$. We note, however, that the lower velocity in the vicinity of the vertices means that a constant $|{\bf l}|$ translates to a reduction in $\tau^{-1}$ on moving away from the nodes, which is contrary to experimental results in the copper oxide materials~\cite{narduzzo1,analytis1,lin1}.

In summary, we have shown that a nodal diamond-shaped pocket with concave sides, produced by charge ordering in the underdoped hole-doped cuprates, provides a natural explanation for the observed sign reversal in the Hall coefficients of YBa$_2$Cu$_3$O$_{6+x}$, YBa$_2$Cu$_4$O$_8$ and HgBa$_2$CuO$_{4+\delta}$~\cite{leboeuf1,doiron3}. Such a diamond-shaped pocket also produces magnetoresistance in the normal state and quantum oscillations in the Hall coefficient~\cite{doiron1}, without the need to invoke multiple Fermi surface pockets~\cite{doiron2,rourke1,chakravarty1}. 

This work is supported by 
the US Department of Energy BES ``Science at 100 T" grant no. LANLF100, the National Science Foundation and the State of Florida. S.E.S.
acknowledges support from the Royal Society, the Winton Programme for
the Physics of Sustainability, and the European Research Council grant
number FP/2007-2013/ERC Grant Agreement number 337425.


\begin{thebibliography}{99}





\bibitem{harris1} J.~M.~Harrison {\it et al.}, Phys. Rev. Lett. {\bf 73}, 1711 (1994).

\bibitem{adachi} T.~Adachi, T.~Noji, Y.~Koike, Phys. Rev. B {\bf 64}, 144524 (2001).

\bibitem{lin1} J.~Lin, A.~J.~Millis, Phys. Rev. B {\bf 72}, 214506 (2005). 

\bibitem{leboeuf1} D.~LeBoeuf {\it et al.}, Nature {\bf 450}, 533 (2007).

\bibitem{doiron3} N.~Doiron-Leyraud {\it et al.}, Phys. Rev. X {\bf 3}, 021019 (2013).

\bibitem{doiron1} 
N.~Doiron-Leyraud {\it et al.} 
Nature {\bf 447}, 565 (2007).

\bibitem{yelland1} E. A.~Yelland {\it et al.}, 
Phys. Rev. Lett. {\bf 100}, 047003 (2008). 

\bibitem{bangura1} A. F.~Bangura~{\it et al.}, 
Phys. Rev. Lett. {\bf 100}, 047004 (2008).

\bibitem{barisic1} N. Bari\u{s}i\'{c} {\it et al.}~{\it et al.}, 
Nature Phys. {\bf 9}, 761-764 (2013).

\bibitem{andersen1} O.~K.~Andersen {\it et al.}, J.~Phys. Chem. Solids~{\bf 56}, 1573 (1995).


\bibitem{millis1} A.~J.~Millis, M.~R.~Norman, Phys. Rev. B~{\bf 76}, 220503 (2007).

\bibitem{chakravarty1} S.~Chakravarty, H.-Y.~Kee, Proc. Natl. Acad. Sci. USA~{\bf 105}, 8835 (2008).

\bibitem{yao1} H.~Yao, D.~H.~Lee, S.~A.~Kivelson, Phys. Rev. B~{\bf 84}, 012507 (2011).

\bibitem{harrison1} N.~Harrison, S. E. Sebastian, Phys. Rev. Lett. {\bf 106}, 226402 (2011).

\bibitem{wu1} T.~Wu {\it et al.}, 
Nature {\bf 477}, 191 (2011).

\bibitem{ghiringhelli1} G.~Ghiringhelli {\it et al.}, Science {\bf 337}, 821 (2012).

\bibitem{chang1} J.~Chang {\it et al.}, Nature Phys.{\bf 8}, 871 (2012).


\bibitem{sebastian1} S. E. Sebastian, S.~E.~{\it et al.}, Nature {\bf 511}, 61-64 (2014).

\bibitem{sebastian9} S.~E.~Sebastian, N.~Harrison, G.~G.~Lonzarich, Rep. Prog. Phys. {\bf 75}, 102501 (2012).

\bibitem{allais1} A.~Allais, D.~Chowdhury, S.~Sachdev, 
Nature Comm. DOI: 10.1038/ncomms6771 (2014).

\bibitem{zhang1} L.~Zhang, J.-W. Mei,  
preprint arXiv:1408.6592 (2014).

\bibitem{tabis1} W. Tabis~{\it et al.}, 
preprint arXiv:1404.7658 (2014).

\bibitem{doiron2} N.~Doiron-Leyraud~{\it et al.}, Nature Comm. 
DOI: 10.1038/ncomms7034 (2015).

\bibitem{senthil1} T.~Senthil, 
preprint arXiv:1410.2096 (2014).

\bibitem{leboeuf2} D.~LeBoeuf {\it et al.}, Phys. Rev. B {\bf 83}, 054506 (2011).

\bibitem{rourke1} P.~M.~C.~Rourke {\it et al.}, Phys. Rev. B {\bf 82}, 020514(R) (2010).

\bibitem{banik1} N.~C.~Banik, A.~W.~Overhauser, Phys. Rev. B {\bf 18}, 1521 (1978).



\bibitem{shockley1} W.~Shockley, Phys. Rev. {\bf 79} 191 (1950).

\bibitem{chambers1} R.~G.~Chambers, Proc. Roy. Soc. London Section 
A {\bf 65}, 458 (1952).

\bibitem{jones1} H.~Jones, C.~Zener, Proc. Roy. Soc. A {\bf 145} (1934).

\bibitem{shoenberg1} D.~Schoenberg, {\it Magnetic oscillations in metals}, (Cambridge University Press, Cambridge 1984).

\bibitem{ong1} N.~P.~Ong, Phys. Rev. B {\bf 43}, 193 (1991).

\bibitem{pippard1} A. B. Pippard, {\it Magnetoresistance in Metals} (Cambridge
University Press, Cambridge 1989).

\bibitem{kikugawa1} N.~Kikigawa {\it et al}, J. Phys. Soc. Japan {\bf 79}, 024704 (2010).

\bibitem{riggs1} S.~C.~Riggs {\it et al.}, 
Nature Phys. {\bf 7}, 332 (2011).

\bibitem{narduzzo1} A.~Narduzzo {\it et al.}, Phys. Rev. B {\bf 77}, 220502(R) (2008).

\bibitem{analytis1} J.~G.~Analytis {\it et al.}, Phys. Rev. B {\bf 76}, 104523 (2007).



%
%
%
%
%
%
%
%
%
%
%
%
%
%
%
%
%
%
%
%
%
%
%
%
%
%
%
%
%
%
%
%
%
%
%
%
%
%
%
%
%
%
%
%
%
%
%
%
%
%
%
%
%
%
%
%
%
%
%
%
%
%
%
%
%
%
%
%
%
%
%
%
%
%
%

%
%
%
%
%
%
%
%
%
%
%
%
%

%
%
%
%
%
%
%
%
%
%
%
%
%
%

%
%
%
%
%
%
%
%
%
%
%
%
%
%
%
%
%
%
%
%
%




%
%
%
%
%
%
%
%
%
%
%
%
%
%
%
%
%
%
%
%
%
%
%
%
%
%
%
%
%
%
%
%
%
%





%
%
%


%
%
%
%
%
%

%
%
%
%
%
%
%
%
%
%
%
%
%
%
%
%
%
%
%
%
%
%
%
%
%
%

%
%
%
%
%
%
%
%
%
%
%
%
%










\end{thebibliography}
\end{document}